# A PHYSICS-BASED LIFE PREDICTION METHODOLOGY FOR THERMAL BARRIER COATING SYSTEMS

E P Busso*, L Wright•, H E Evans†, L N McCartney•,
S R J Saunders•, S Osgerby• and J Nunn•

\* Centre des Matériaux, Ecole des Mines de Paris, CNRS UMR 7633, Evry, FRANCE
† Department of Metallurgy and Materials, University of Birmingham, UK
• National Physical Laboratory, UK

**ABSTRACT**

A novel mechanistic approach is proposed to predict the life of thermal barrier coating (TBC) systems. The life prediction methodology is based on a criterion linked directly to the dominant failure mechanism. It relies on a statistical treatment of the TBC's morphological characteristics, non-destructive stress measurements and on a continuum mechanics framework to quantify the stresses that promote the nucleation and growth of microcracks within the TBC. The latter accounts for the effects of TBC constituents' elasto-visco-plastic properties, the stiffening of the ceramic due to sintering and the oxidation at the interface between the thermally insulating yttria stabilized zirconia layer and the metallic bond coat. The mechanistic approach is used to investigate the effects on TBC life of the properties and morphology of the top YSZ coating, metallic low-pressure plasma sprayed bond coat and the thermally grown oxide. Its calibration is based on TBC damage inferred from non-destructive fluorescence measurements using piezo-spectroscopy and on the numerically predicted local TBC stresses responsible for the initiation of such damage. The potential applicability of the methodology to other types of TBC coatings and thermal loading conditions is also discussed.

# 1. INTRODUCTION

The ability to predict the lifetime of thermal barrier coating (TBC) systems is a high priority for gas turbine users. Typical TBC systems are made of a top layer of yttria stabilized zirconia (YSZ -the insulating layer) and a metallic low pressure plasma sprayed (LPPS) bond coat. During exposure at high temperatures, the metallic bond coat forms a thermally grown oxide (TGO) that consists predominantly of alumina. The formation of the TGO plays a crucial role in the failure of the TBC. The usual mechanism of TBC failure is associated with spallation at or close to the TGO interfaces within either the YSZ or the bond coat. A typical crack pattern in an EB-PVD MCrAlY bond coat after 1000 $^{o}$C exposure for 700 h is shown in Fig. 1(a). In the micrograph of Fig.1(b), possible initiation sites are also shown of interfacial microcracks between the TGO and the bond coat of an EB-PVD TBC system exposed to 1000 $^{o}$C for 600 h.

Approaches to predict the mechanical integrity of TBCs have generally concentrated on the development of life prediction models based on empirical/phenomenological fatigue life relationships which link the life of the TBC to the summation of damage due to mechanical straining and oxidation [1]. Delamination models, which have been considered in some theoretical depth, deal with edge cracking and buckling [2]. However, the latter are widely perceived only as a possible but not critical part of the failure process since an extensive delamination zone at the base of the ceramic top coat is necessary for crack propagation and buckling to occur. The sub-critical growth of such a zone is generally the slow, life-determining stage in TBC systems. However, the conditions under which this could occur have received little attention.

In this work, a novel mechanistic approach is proposed to predict the life of TBC systems. It relies on a statistical treatment of the TBC's morphological characteristics and on a continuum mechanics framework to quantify the local TBC stresses responsible for sub-critical crack nucleation and growth. Two key aspects characterise this life prediction methodology. Firstly, it incorporates the complex interaction between interfacial and microstructural features, local oxidation mechanisms and time-dependent processes (i.e. oxidation kinetics, diffusion, creep and sintering of the ceramic material). Secondly, its calibration is based on TBC damage inferred from non-destructive measurements and on the numerically predicted local TBC stresses responsible for the initiation of such damage. The non-destructive measurements are based on a combination of fluorescence measurements using piezo-spectroscopy and high resolution thermal images. Thus, the methodology requires TBC damage information generated by non-destructive measurements, a suitably defined failure criterion based on the dominant failure mechanisms, morphological characteristics of the TGO obtained from microstructural observations, and a numerically generated database containing the local predicted TBC stresses.

The emphasis of the paper will be on the different stages involved in the proposed life prediction methodology and on how to integrate the experimental and numerical data. Moreover, the mechanistic study of local TBC stresses will be summarised and, when necessary, reference to related published work will be given where further details about the approach can be found. The full validation of the methodology and an investigation into the effects of significant thermal cycling are to be reported in a future publication. The paper is structured as follows. In Section 2, an overview of the TBC damage measurements and overall mechanistic approach is first given. Representative numerical results of the TBC stress analyses are then presented in Section 3. Section 4 describes the different stages involved in the TBC life prediction methodology and highlights its predictive capabilities and potential for further development. Finally, Section 5 presents some general conclusions.

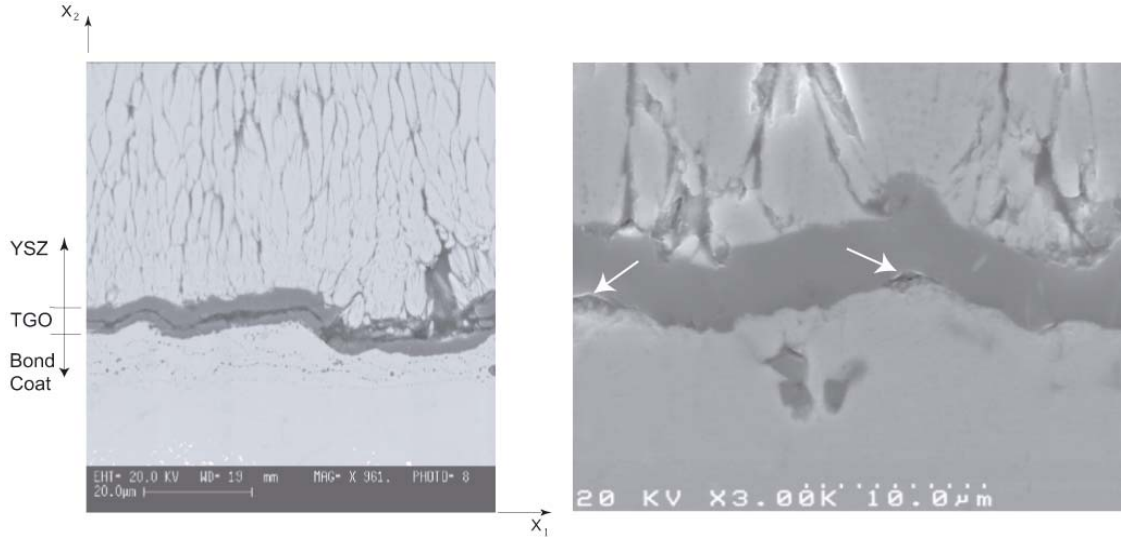

**Figure 1** SEM micrographs showing (a) a typical microstructure of a TBC system (2466 MCrAlY bond coat, EB-PVD YSZ, CMSX4 substrate) after 700 h at 1000 $^{o}$C, and (b) possible initiation sites (see arrows) of interfacial microcracks between the TGO and the bond coat of a TBC system (bond coat: 2412 MCrAlY, EB-PVD YSZ, substrate: CMSX4) exposed to 1000 $^{o}$C for 600 h.

## 2  Mechanistic Approach To Model TBC Stresses

### *2.1  Mechanistic Considerations*

Experimental work (e.g. [1-5]) has shown that the failure modes of EB-PVD TBCs are very dependent on the properties of the bond coat, the YSZ, and the TGO growth rate. In addition, the stress conditions in and around the TGO layer are significantly affected by the aspect ratio of the wavy TGO morphology. One of the most commonly observed TBC failure mechanisms is the spallation of the top YSZ coating driven by the high interfacial tractions that develop at the TGO interfaces upon cooling (e.g. see [3-5]). Experimental evidence of interfacial damage is shown in Fig. 1(b), where the arrows indicate possible initiation sites of interfacial cracks between the TGO and the bond coat. The driving force for the nucleation of such types of defects is intrinsically linked to the local stress conditions, as illustrated in Fig. 2. Here, a schematic diagram of a TBC system shows the different relevant stresses and tractions, which need to be considered when studying the failure of the rough undulating interfaces often found in TBC systems. The out-of-plane stress component, parallel to the $X_2$ axis in Fig. 2, is responsible for the formation of microcracks lying in the $X_1$-$X_3$ plane, which are generally seen within the YSZ near the TGO interface. The interfacial TGO tractions, normal to either the top coat-TGO or the bond coat-TGO interfaces, drive the nucleation of interfacial microcracks and propagation of interfacial defects. Finally, the maximum principal stress is generally identified as the main 'driving force' for the nucleation of cohesive microcracks within the TGO.

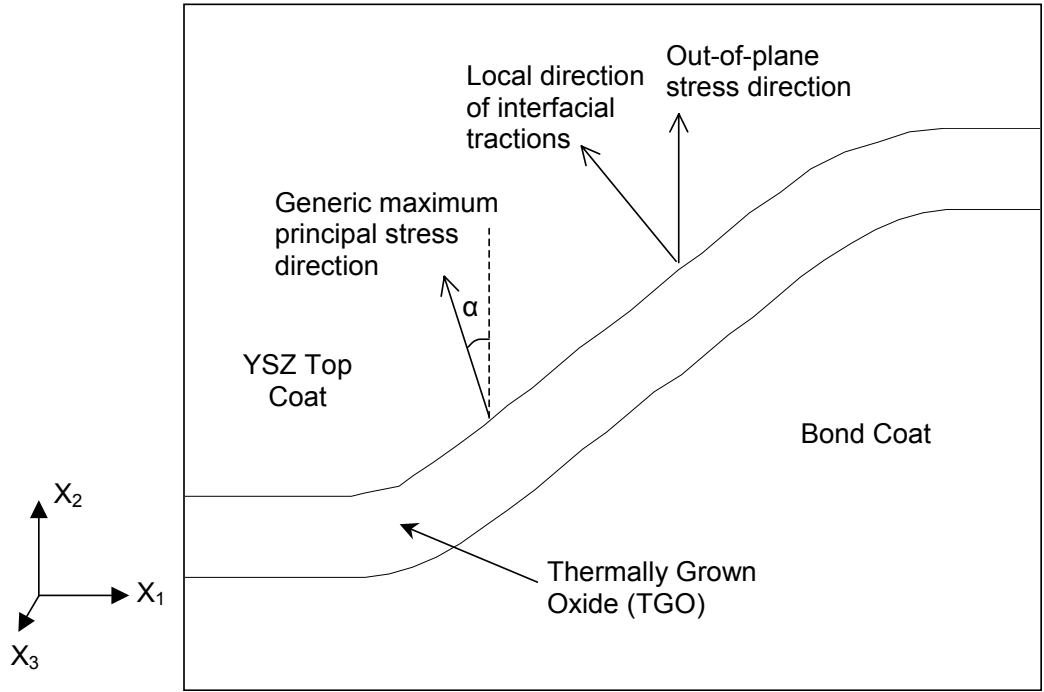

**Figure 2** Diagram illustrating the structure of the TBC system and some of the stresses that are used in determining the failure of the system.

### 2.2  TBC Damage Measurements

The non-destructive measurements of TBC damage were based on a combination of high resolution thermal images and fluorescence measurements using Raman spectroscopy. The latter relies on the fluorescence of trace amounts of Cr ions in the TGO and on the fact that spectral positions depend on the magnitude of the stresses in the sampled TGO volume. Thus, stress changes associated with microcracking / debonding of the alumina layer can be readily detected. The stress state can be estimated from the peak shift of the spectra with respect to the unstressed sapphire reference. In the most general case, the frequency change is related to the stress components via the relation

$$\Delta \nu = \Pi_{11}\sigma_{11} + \Pi_{22}\sigma_{22} + \Pi_{33}\sigma_{33} \tag{1}$$

where $\Pi_{ij}$ are piezo-spectroscopic coefficients for the system and $\sigma_{ij}$ are the stress components. By assuming a random distribution of the crystallographic texture in the TGO, isotropic properties apply so that $\Pi_{11} = \Pi_{22} = \Pi_{33} = \Pi$ leading to

$$\Delta \nu = \Pi(\sigma_{11} + \sigma_{22} + \sigma_{33}) , \tag{2}$$

implying that what is measured is the hydrostatic stress, an invariant of the stress field. Alternatively, if an equi-biaxial plane stress state is assumed, then the frequency shift can be related to just the biaxial stress component (e.g., see Selculk and Atkinson [6]). However, such a plane stress assumption is particularly suited to rather flat oxide layers which, as will be discussed later on in the text, are not generally seen in TBC systems. Such an approximation will not be correct for rough TGOs, as the waviness of the TGO gives rise to non-zero stress components in the direction normal to the TBC surface.

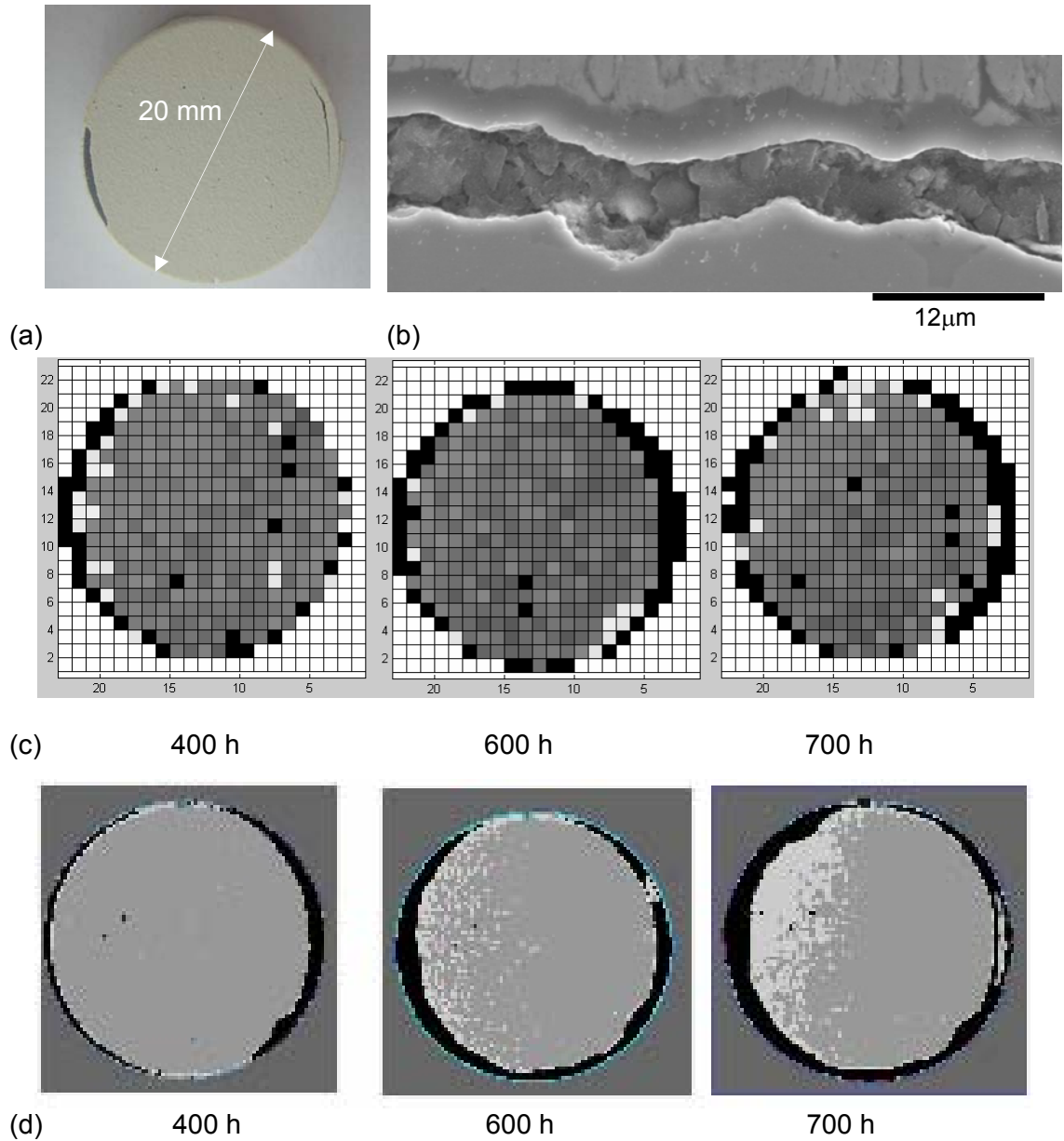

**Figure 3** Typical damaged area maps measured in the current TBC system (2466 MCrAlY bond coat, EBPVD YSZ on a CMSX4 substrate): (a) photograph of a coated 20 mm diameter test sample after 700 h test at 1000 $^0$C, (b) typical SEM micrograph of a failing TBC system (mean TGO thickness is 3.3 μm), (c) stress maps measured by piezo-spectroscopy and (d) corresponding thermographs after 400, 600 and 700 hours exposure at 1000 $^0$C. (The sampling points in (c) were taken at sites in an array of cell dimensions 1x1 mm.)

Typical results from the experimental measurements of TBC damage accumulation with thermal loading, conducted in this study to underpin the development of the life prediction methodology, are shown in Figs. 3(a) to (d). Figure 3(a) shows a photograph of a TBC coated superalloy sample oxidised for 700 h at 1000°C. Delaminated areas around the edge of the specimen can be clearly seen. Figure 3(b) shows an SEM micrograph of a typical local TGO crack, which appears to have propagated along the TGO-bond coat interface. In Fig. 3(c), the results of TBC damage measured by piezo-spectroscopy are shown at three different oxidation times. Here, the grey area represents a compressive stress of around 4 GPa and black regions represent 0 GPa. Thus, the red areas are measured regions where the stresses have been relaxed by damage (or microcracks). The identification of such regions requires the specification of a threshold value for the measured mean stress to define the boundary between the 'damaged' and 'undamaged' parts of the TBC. The required damaged region is defined as a region initially under high compressive stress that moves to a lower compressive or tensile stress state such that the mean stress is above the defined threshold value. The results obtained by piezo-spectroscopy can be corroborated metallographically by examining sections taken through the TBC sample in order to define a relationship between observed cracking/damage and the measured mean stress in the damaged region. Alternatively, thermography may also provide the required information (see for example [7]), as shown by the thermography measurements of Fig. 3(d) conducted on the same specimens as those used for the piezo-spectroscopy measurements. The thermographs shown in Fig. 3(d) were obtained at room temperature when radiating the surface of the specimen with low level infra-red. The apparent temperature difference between the black and grey pixels shown on the specimen in this figure is of the order of 1°C, and the background is approximately 3°C cooler than the specimen. A comparison between Figs. 3(c) and (d) reveals good agreement between the stress-relieved areas with the coolest ones from the thermal images, giving credence to the understanding that these are indeed areas where damage has occurred.

## 2.3 Mechanistic Approach to Determine the Local TBC Stresses

The mechanistic approach followed here is based on the work of Busso *et al*. [3-5] and accounts for the effects of YSZ sintering, bond coat creep and TGO morphology on the local stresses responsible for interfacial cracking. The oxidation-induced stresses that control the TBC degradation and eventual failure, such as the components illustrated in Fig. 2, are calculated using a constitutive framework coupled with the diffusion of oxidant species through a multi-phase solid. The calculations are performed using the ABAQUS finite element (FE) analysis code. The overall thickness of the oxide layer is taken to be a phenomenological function of temperature and time. The proposed internal state variable constitutive model accounts for the phase transformation associated with the oxidising phase, and incorporates the effect that the local volumetric expansion of the newly formed oxide has on the generation of inelastic volumetric strains and stresses.

The strains associated with the oxidation-induced phase transformation of the bond coat into the internal primary oxide were calculated in several stages:

1. The total thickness $h$ of the oxide layer at the current time was first computed from a phenomenological oxide growth kinetics relation in terms of temperature and accumulated time, t, at the maximum temperature in the cycle, $\theta_{max}$ [5],

$$h = A_0 \, t^q \exp\left(\frac{Q_O}{R}\left(\frac{1}{\theta_R} - \frac{1}{\theta_{max}}\right)\right) \qquad (3)$$

   where $A_0$ is a proportionality constant, $q$ a growth exponent, $Q_0$ an apparent activation energy, $\theta_R$ a reference temperature, and $R$ the universal gas constant. For the TBC system of interest in this work, the calibrated oxidation parameters are $A_0 = 10^{-6}$ m/s$^q$, $\theta_R = 2424$ K, $Q_0 = 766{,}900$ J/mol and $q = 0.332$. Thus the units of h and $\theta_{max}$ are m and K, respectively.

2. The total oxide thickness was then subdivided into an internally grown part, of thickness, $h_{in}$, and an externally grown component of thickness, $h_{out} = h - h_{in}$, with their ratio given by $h_{in}/h_{out} = 1/(PBR - 1)$. Here PBR = 1.28 is the Pilling-Bedworth ratio corresponding to the primary oxidation reaction.

3. The volumetric strains associated with the oxidation reaction, whose mean value is given by $e^T = \log_e(PBR / 3)$, were applied to the row of elements in the FE model undergoing internal oxidation. Note that this expansion is irreversible and generally anisotropic, with the anisotropy of the transformation determined experimentally. The in-plane normal strains are given by $\varepsilon_{X1} = \varepsilon_{X3} = 3\ e^T\ r/(1 + 2r)$, and the out-of-plane normal strain is $\varepsilon_{X2} = 3\ e^T\ /(1 + 2r)$, where $r = 0.005/0.435 = 0.0115$ (see [5] for further details).

The constitutive model for the oxidising bond coat material was implemented into the FE method as a user-defined material subroutine and used to update the stresses at each material point within the unoxidised and oxidising bond coat in the finite element models of the TBC, as described in the next Section. For further details about the approach, refer to [3-5].

The temperature dependence of the elastic material properties associated with the bond coat and the TGO are shown in Table A1 of Appendix A. The table also includes the yield stress of the bond coat, used in an incremental model of perfect plasticity, which is separate from the power law creep model used for the bond coat and the TGO (see Appendix A). The YSZ top coat is modelled as a transversely isotropic elastic solid, with properties dependent on the degree of sintering (parameterised by sintering time) and temperature. The YSZ elastic moduli, taken from [5], are given in Appendix A.

## 2.4 Identification of an Appropriate Criterion for TBC failure

It is expected that, for a given oxidation temperature, a TBC system of the type studied here would fail by the time the oxide reaches a critical thickness. The definition of the critical TGO thickness can be made from the type of data presented in Fig. 4, which shows the effect of oxidising temperature on the time to TBC spallation for MCrAlY type coatings and for PtAl based TBCs. The data points in the figure taken from the open literature [8] are the closed symbols. The open symbols represent the data obtained as part of this work. In each case, circles and diamonds are MCrAlY coatings and triangles are PtAl. As in the original data set, there does not appear to be a systematic difference between the two sets although the current data set on the aluminides is limited. The striking feature of these data is the large spread (a factor of ~10) in the times to spallation at any given temperature, even for the present samples which were nominally identical and tested in the same facility. This large scatter may be due to an uncontrolled variation in the bond coat surface roughness [8] since this is expected to produce a corresponding variation in the stresses developed within the TBC system. An objective of the present FE analysis is to evaluate this possibility. The results from the current data, in this regard, are typical of those found in the literature. They also show similar failure times to those reported elsewhere so that the two sets of data appear mutually consistent. The continuous line in Fig. 5 represents the time at a given temperature to reach an alumina TGO thickness of 3 μm, calculated using Eq. 3. It can be seen that this prediction at the 3μm TGO thickness level forms an approximate lower bound to the data set with the possible exception of tests at 1000 °C where they provide an underestimate.

## 2.5 Identification of Representative Periodic Unit Cells and FE Model

The morphological characteristics of the TGO interface regions of the EB-PVD MCrAlY bond coating, see Figs. 1 and 2, can be represented by the series of undulations shown in Fig. 5. Here, the random nature of the interface morphology can be idealised by a periodic representative profile such as the unit cell shown in Fig. 5. Two interface roughness parameters characterise the unit cell, namely (i) the half-period *a* and (ii) the amplitude *b*. Note that whilst *b/a* constitutes the most important roughness parameter, *b* is also required to account for the fact that its value relative to that of the TGO thickness will affect the local stresses.

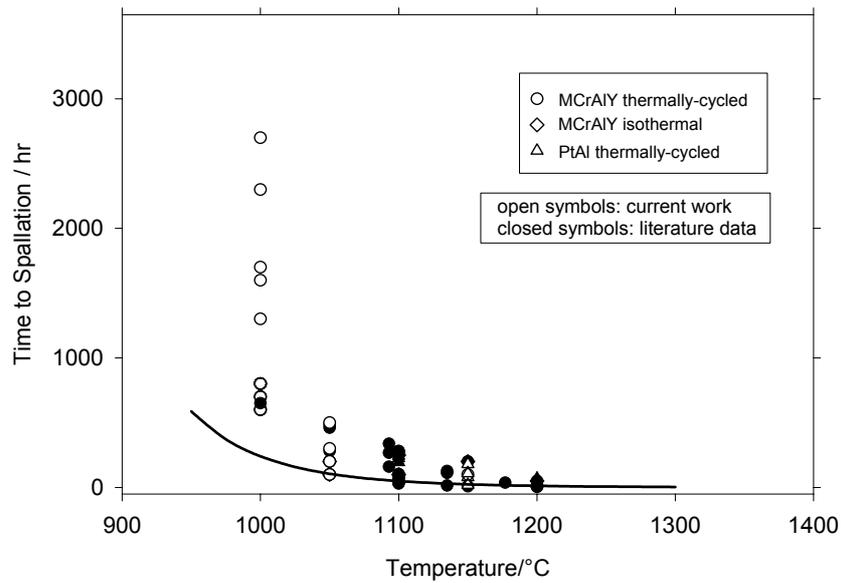

**Figure 4** Experimental data showing the effect of oxidising temperature on the time to TBC spallation for MCrAlY type coatings and for PtAl based TBCs (triangles). The line represents the time at a given temperature to reach a TGO thickness of 3 $\mu$m, calculated using Eq. 3.

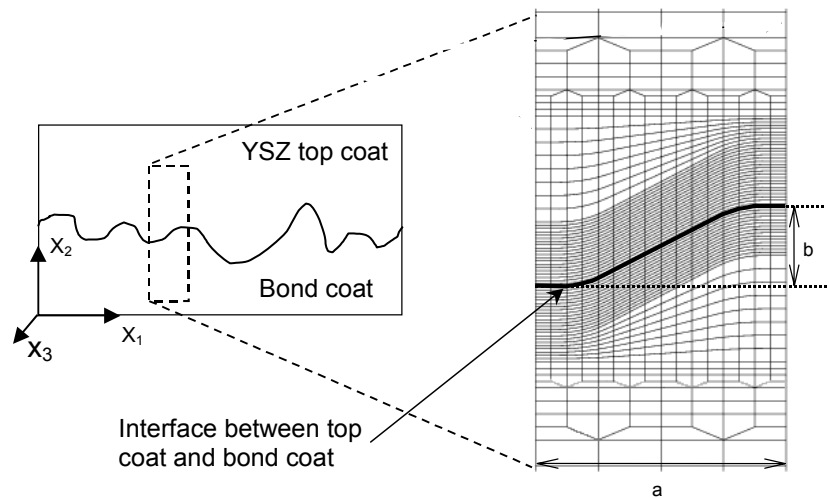

**Figure 5** Illustration of the surface roughness geometric parameters *a* and *b* and a typical finite element mesh of a periodic unit cell used in the modelling.

Typical TGO-YSZ profiles were identified from SEM micrographs of the interfacial regions for the EB-PVD TBC /LPPS bond coat system of interest and analysed to obtain the statistical distributions of the geometric parameters *b* and *b/a*. In order to obtain the representative spread of '*b*', '*a*' and '*b/a*' ratios of the undulations of the bond coat surface, a specimen was cross sectioned and polished to reveal the interface. Over 100 measurements were made on a total of 28 scaled micrographs. Typical mean and standard deviation roughness values for the TBC system were found to be 2.39 $\mu$m and 1.62 $\mu$m for *b* and 0.32 and 0.12 for *b/a*. Three sets of interfacial roughness parameters (b, b/a) were chosen for the stress analysis calculations from the measured roughness distributions, so as to be as close as possible to the mean value ± one standard deviation for each interfacial parameter. This yielded,

- $b$ = 0.79 µm, $b/a$ = 0.20,
- $b$ = 2.47 µm, $b/a$ = 0.32, and                                                                       (4)
- $b$ = 4.54 µm, $b/a$ = 0.48.

The TBC is assumed to be subjected to a generic temperature cycle consisting of three stages: a heating transient from a minimum $\theta_{min}$ to a maximum, $\theta_{max}$, at a constant rate, followed by a dwell period of length, $t_{hold}$, and a cooling transient to $\theta_{min}$ at the same rate as in the heating transient. In the work conducted here, $\theta_{min}$ is always assumed to be 25 °C, and $\theta_{max}$ is set to typical TGO temperatures found during service in land-based gas turbines, namely either 1000 °C or 1100 °C. Based on the experimental evidence shown in Fig. 4 and the discussions in Section 2.4, the holding time $t_{hold}$ was chosen such that the oxide layer would have reached the critical thickness of 3 µm after one thermal cycle.

As previously discussed, the aim of the FE modelling work was to carry out a set of parametric calculations, the results of which are required as inputs for the prediction of coating life, as will be described in Section 4. The parametric variables in the numerical study were the maximum cycle temperature, $\theta_{max}$, and the interfacial parameters $b/a$ and $b$ (here assumed to be perfectly correlated). Finite element models of the periodic unit cells for the three representative TGO-YSZ profiles were then developed. The models of the TGO-YSZ interfacial regions were constructed from the single geometric unit of the form shown in Fig. 5. The surface of the TBC was assumed to be traction free, and symmetry and periodic boundary conditions were assumed on the planes $X_1$ = 0 and $X_1$ = $a$, respectively (see Fig. 5). In this way, the interface is effectively modelled as an infinite series of undulations of period $2a$ and amplitude $b$. The initial thicknesses of the bond coat and YSZ top coat were taken to be 50 and 125 µm, respectively. The temperature within the TBC system was assumed to be uniform.

The 2D model of the TBC system relied on generalised plane strain assumptions (i.e. strain along $X_3$, see Fig. 5, is uniform but not necessarily zero). Furthermore, the TBC system was assumed to be attached to a bulk substrate whose deformation is determined solely by its thermal expansion, that is, it is elastically rigid so that its Young's and shear moduli are effectively assumed to be infinite. This implies that the in-plane displacements of the $X_1$ and $X_3$ boundaries of the representative unit cell model are controlled by the expansion and contraction of the substrate. Note that the shape of the curved parts of the periodic unit cell is chosen to be consistent with typical TGO interface profiles observed in SEM micrographs such as that shown in Fig. 1. Here, the interface morphology was constructed by linking straight line segments and circular arcs in order to yield a curve with a continuous first derivative.

## 3. Stresses Obtained from Parametric FE Studies

Some representative results from the finite element analyses are presented and discussed in this Section.

Figures 6 (a) to (c) shows the contour plots of the out-of-plane stress component ($\sigma_{22}$) when the TBC was cooled down to 25 °C after 241 h oxidation at 1000 °C. The results are shown for TBCs with the three different TGO morphologies specified in Eq. 4. Note that the TGO thickness is 3 µm in all the plots. Similar results were obtained for oxidation at 1100 °C, but contours at this higher temperature are not shown in this paper for brevity. It can be seen that the extent and magnitude of the most highly stressed areas increases with interface roughness, and that these regions are located in the YSZ top coat around the valleys. It is also worth noticing that both the TGO and the YSZ regions tend to be generally under some compression at 25 °C for the rougher TBCs. Thus these results show that, for smoother interfaces, peak tensile stresses develop within the TGO but, for rougher interfaces (such as those shown in Fig. 1), significant tensile stresses occur at the bond coat / TGO apex.

Figures 7(a) to (c) show the contour plots of the out-of-plane stress component $\sigma_{22}$ when the TBC was cooled down to 25 °C after (a) 61 h, (b) 167 h and (b) 241 h oxidation at 1000 °C. At this temperature, the corresponding oxide thicknesses are 1.91, 2.65 and 3.00 μm, respectively. The initial interfacial geometry used was $b$=2.47 μm, $b/a$ = 0.323. It can be seen that the most highly stressed regions are in the YSZ around the valleys, and that the magnitudes of the out-of-plane stresses in these regions increase rapidly with oxide thickness. Results at 1100 °C show similar trends. In general, a comparison between the results obtained at different temperatures shows clearly that the TBC stresses do not depend just on the oxide thickness but also on the kinetics of oxidation. Whilst the stress contours calculated at different temperatures share some features, there are significant differences.

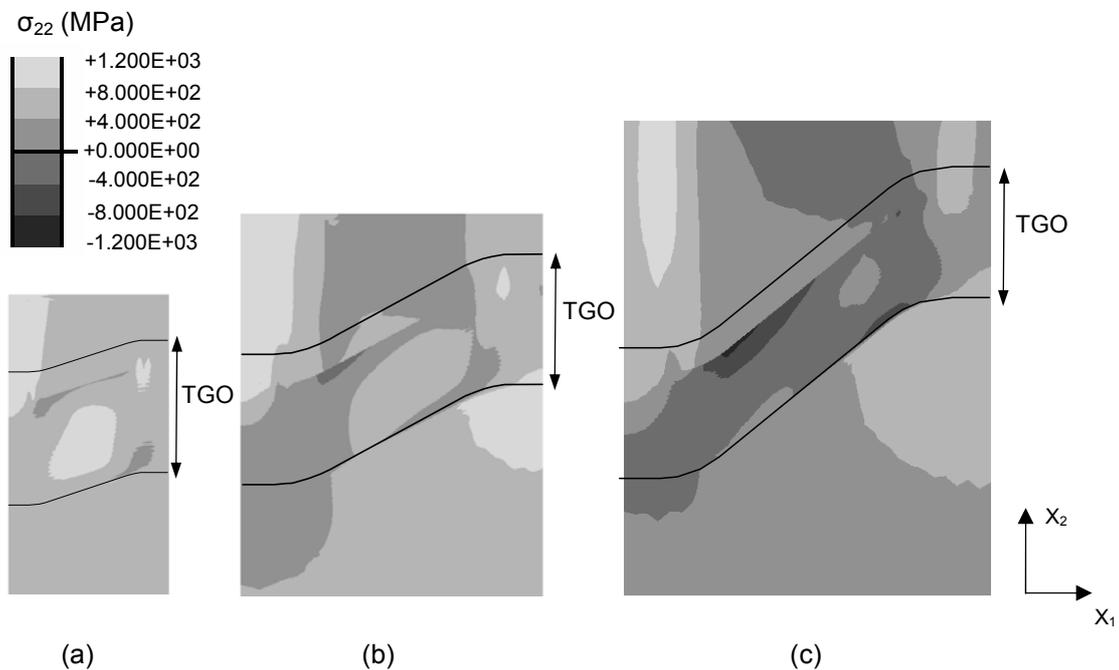

**Figure 6** Three typical out-of-plane stress contour patterns at 25 °C for different geometric parameters with $\theta_{max}$= 1000 °C: (a) $b$ =0.79 μm, $b/a$ = 0.20, (b) $b$= 2.47 μm, $b/a$ = 0.32 and (c) $b$ = 4.54 μm, $b/a$ = 0.48. In all cases, the TGO thickness is 3 μm.

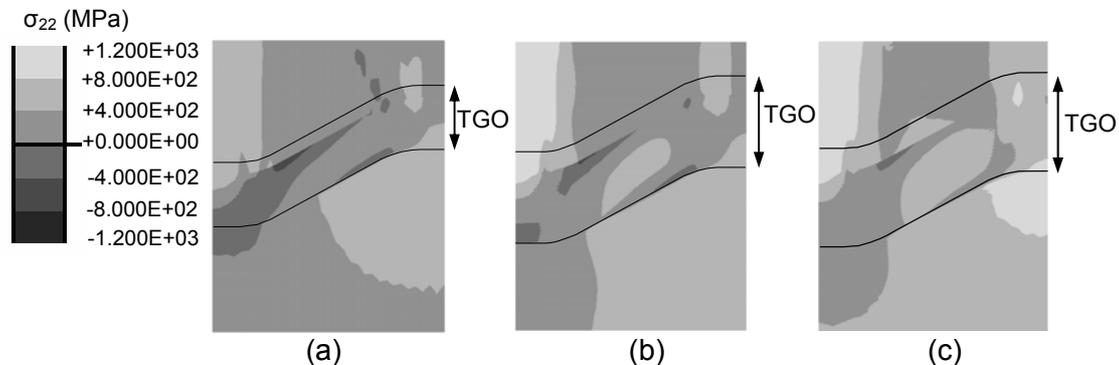

**Figure 7** Contour plots of the out-of-plane stress component $\sigma_{22}$ after cooling to room temperature from 1000 °C for (a) 61 h, (b) 167 h, and (c) 241 h oxidation. The corresponding TGO thicknesses are 1.91, 2.65 and 3.00 μm, respectively. (Also, $b$ = 2.47 μm, $b/a$ = 0.323).

A summary of the results obtained from the parametric finite element analyses at both 1000 and 1100 °C is shown in Appendix B. The predicted effects can be seen of the geometric parameters and peak cycle temperature on the stress maxima relevant to the different TBC failure modes at room temperature.

## 4. Methodology for TBC Life Prediction

The proposed TBC lifetime prediction methodology can be considered as a three-stage approach which requires the following inputs:

- A well characterised thermal cycle (defined by $\theta_{max}$, $\theta_{min}$, and $t_{hold}$, and a specified heating and cooling rate, see Section 2.5).
- The accumulated time defined as $t_{acc} = t_{hold}N$ where $N$ is the number of thermal cycles.
- The dominant failure mechanism for the TBC coating system and loading conditions of interest (as discussed in Sections 2.1 and 2.4).
- A statistical distribution of the TBC's interface roughness (defined by the ratio $b/a$ and amplitude $b$) obtained from actual TBC micrographs.
- Experimental measurements of TBC damage accumulation with thermal loading (see Section 2.2).
- A full description of the material properties of all constituents of the TBC (see Appendix A).
- A numerically generated database of the relevant local TBC stresses obtained from a parametric study such as that presented in Section 3 (see also refs. [3-5]).

A PC-based software tool, using Excel spreadsheets, has been developed to carry out some of the calculations required at the different stages of the life prediction methodology, which are described next. Fig. 8 illustrates the methodology as a flow chart.

Note that all interpolations mentioned in this Section are linear interpolations, unless specified otherwise. Furthermore, here the times have been converted into equivalent thicknesses of TGO and the interpolation has then been carried out with respect to TGO thickness. This enables the interpolation to be carried out more accurately due to the fact that the stresses are likely to depend more linearly on TGO thickness than on accumulated time.

This Section illustrates the methodology using real measurement data and the FE results given in Appendix B. It should be noted that whilst the measured and modelled TBC systems are similar, they are not identical. No experimental data were available for the modelled system. The differences between the systems make some aspects of the software tool less reliable: in particular, the example shown contains more extrapolation of FE results than is desirable due to the long time over which the damage measurements were taken. It is, however, emphasised that the purpose of this paper is to describe the methodology, and not to validate the approach by comparing predictions with experimental results. For validation of the methodology, the FEA runs would need to be repeated using model properties that correspond more closely with those of the measured system.

### 4.1 Stage 1: Quantification of TBC Damage

This stage deals with the way in which the information about TBC damage (see Section 2.2) is integrated into the life prediction methodology. Two critical assumptions are made in this stage. Firstly, the total damaged area within the TBC is assumed to be directly associated with areas exhibiting detectable stress relief, which constitutes a measure of damage as shown in Fig. 3. The measured fraction of damaged area can be quantified by a generic dimensionless scalar variable or "damage parameter", $D$, and expressed as a function of accumulated time at the oxidising temperature, $t_{acc}$, for given values of $\theta_{max}$ and $\theta_{min}$, so that:

$$D = \hat{D}(\theta_{min}, \theta_{max}, t_{acc}), \qquad \text{where } t_{acc} = t_{hold}N, \qquad (5)$$

and N is the total number of thermal cycles. Thus Eq. 5 relates the accumulated time to the proportion of areas sampled that exhibited damage, which can be derived from measured stress maps such as those shown in Fig. 3(c). The measured $t_{acc}$ vs. $D$ curve for the case where $\theta_{max}$ = 1000 °C is shown in Fig. 9. Note that the TBC failed after approximately 700 h oxidation, failure in this case being taken as damage of more than 22%.

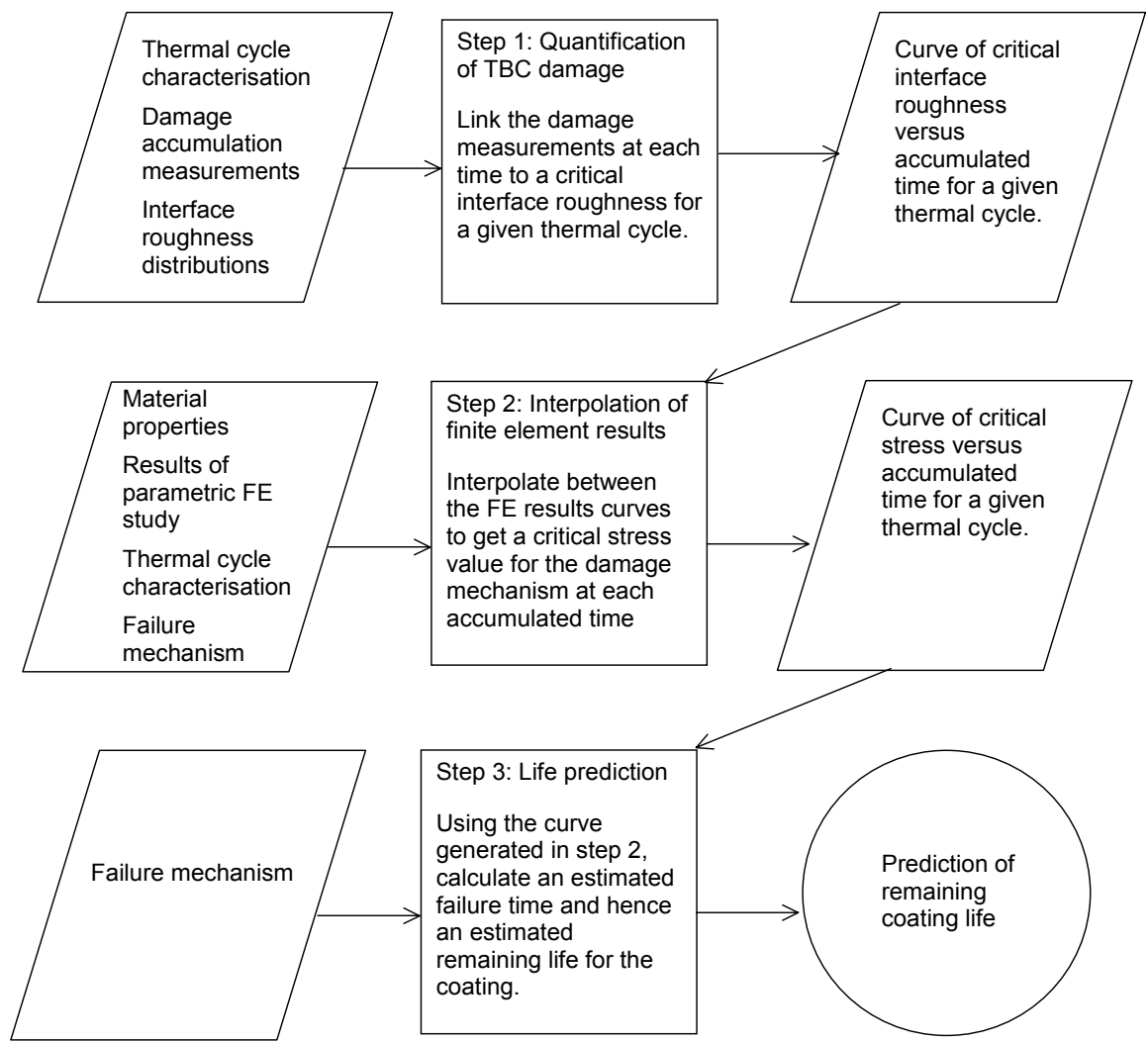

**Figure 8**: Flow chart of the life prediction methodology.

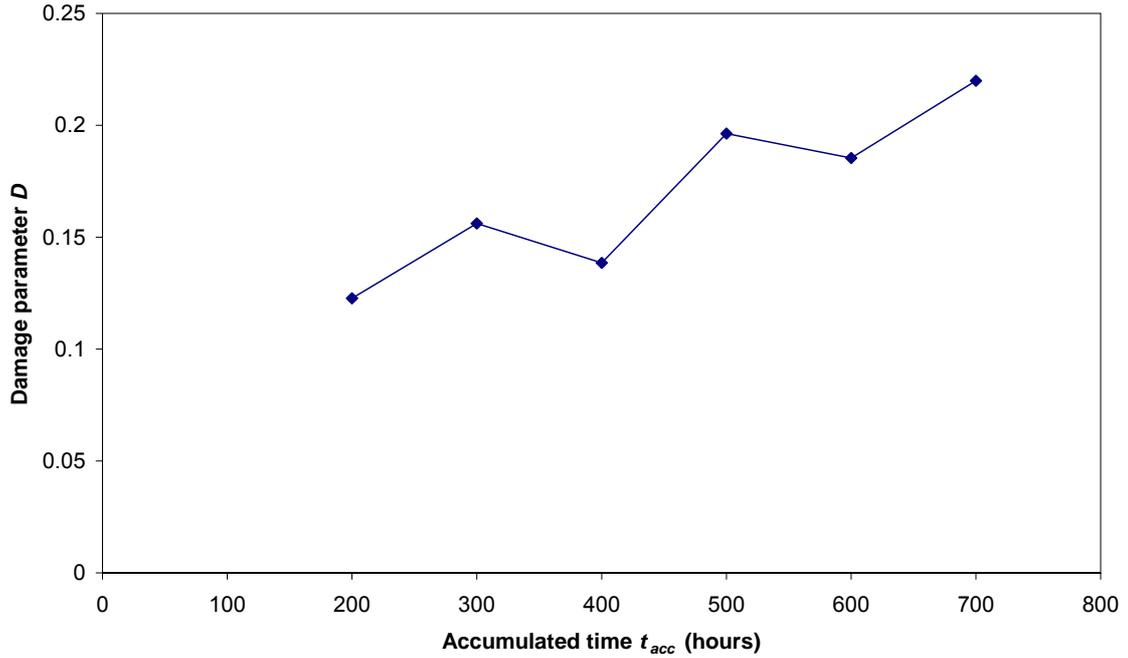

**Figure 9** Measured fraction of damaged area against time, derived from measured stress maps such as those shown in Fig. 3 for the case when $\theta_{max}$ = 1000 °C.

Secondly, this stage assumes that all of the measured damage occurs only at or near interfacial regions having values of *b/a* that are **larger** than some critical value $r_c$. Hence, through the damage measurement *D*, a time $t_{acc}$ can be associated with a critical value $r_c$. This definition of $r_c$ can be written as

$$D = \int_{r_c}^{\infty} f(x)dx, \qquad (6)$$

where *f* is the probability density function of *b/a*. Figure 10 illustrates Eq. 6 graphically, using a fitting of the data obtained from a statistical treatment of the TGO's morphological characteristics. The area under the probability density function of Fig. 10 corresponding to a measured value of *D* when time is equal to $t_{acc}$, enables the estimation of the critical roughness value $r_c$.

A more convenient way of determining the critical value $r_c$ from a given measured value of $D$ is by replotting the information contained in Fig. 10 as the corresponding cumulative probability function. This is shown in Fig. 11, which can now be used directly to obtain the critical b/a values corresponding to every measurement of $D$ that is available, making it possible to assemble a curve of $r_c$ against $t_{acc}$ for the given thermal history. It is worth pointing out that the dependence of Fig. 11 on temperature is only through the use of $t_{acc}$ since the relationship between $D$ and b/a is only a property of the coating and not of the thermal cycle. The symbols in Fig. 11 show the cumulative distribution of the measured data, and the line without symbols shows the normal distribution with the same mean and variance, i.e. the integral of that shown in Fig. 10.

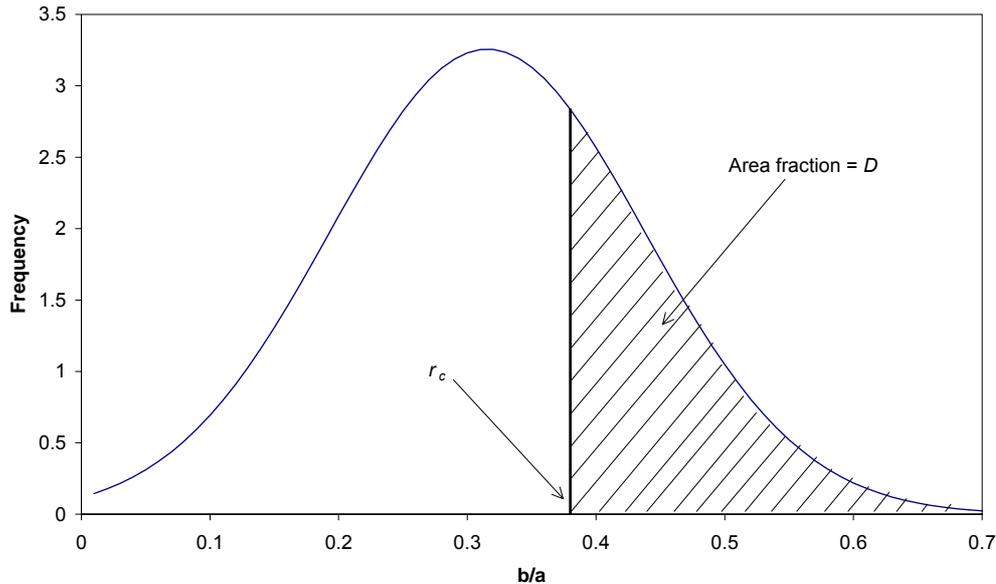

**Figure 10** Critical value $r_c$ as determined from the statistical normal distribution of the b/a values.

It is important to note the differences between using the statistical normal distribution shown in Fig. 10, and the measurement data shown in Fig. 11. It is clear from Fig. 10 that the use of the normal distribution means that there is a non-zero probability of obtaining a negative value of $r_c$. This is physically meaningless since the definitions of $b$ and $a$ are lengths rather than directed vectors. This problem could be removed if a different distribution (such as the log-normal) was used to describe b/a.

The output from this stage is the relationship between the critical interfacial roughness ratio $r_c$ and accumulated oxidation time $t_{acc}$ for a given set of $\theta_{max}$ and $\theta_{min}$ temperatures. This can be obtained by combining the information presented in Figs. 9 and 11. The resulting $t_{acc}$ vs. $r_c$ relation for the thermal cycle with $\theta_{min} = 25\ ^\circ C$ and $\theta_{max} = 1000\ ^\circ C$ is shown in Fig. 12. Thus, this curve gives the minimum value of roughness of TGO regions that will develop damage (i.e. cracks) after the TBC has been exposed to the oxidising temperature (1000 $^\circ$C) for different lengths of (accumulated) time.

If measurements of $D$ generated during different thermal cycles are available, then this step can be repeated many times for other values of $\theta_{max}$ and $\theta_{min}$. If enough curves are generated in this way, then interpolation between the damage versus time curves will enable the expected response at other values of $\theta_{max}$ and $\theta_{min}$ to be estimated.

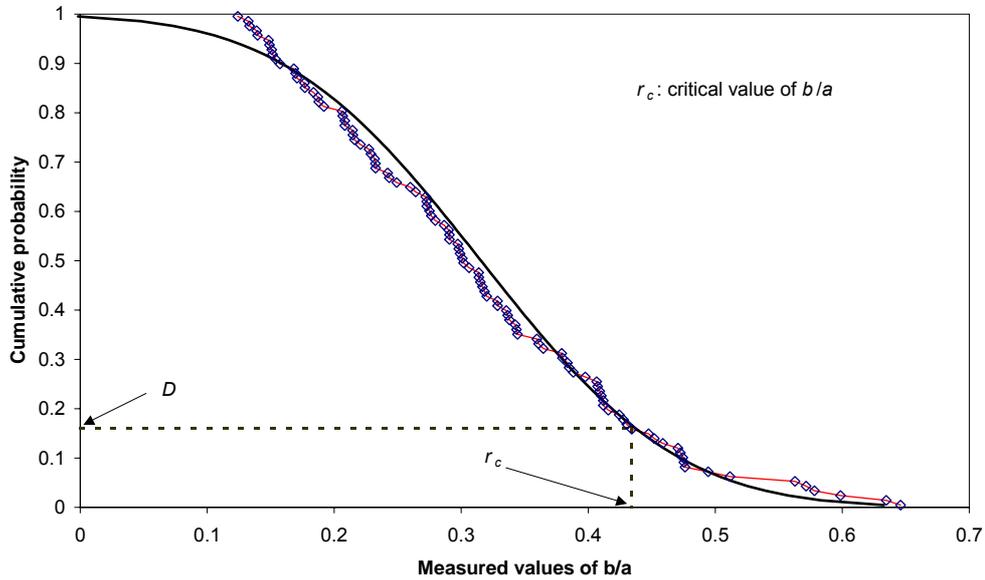

**Figure 11** Determination of the critical value of $b/a$ from its cumulative distribution function. The line with symbols shows the measurement data, and the smooth line without symbols is the cumulative normal distribution with the same mean and variance as the measured data.

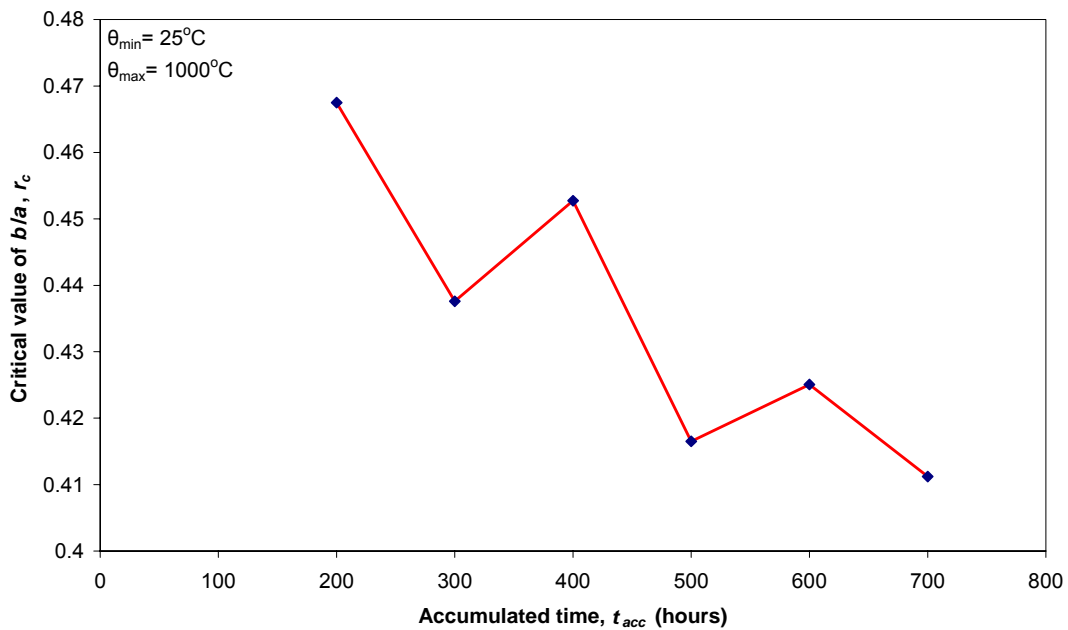

**Figure 12** Critical value of $b/a$, $r_c$, as a function of accumulated time for a given thermal history.

### *4.2    Stage 2: Interpolation and Extrapolation of the Finite Element Results*

This stage combines the $r_c$ vs. $t_{acc}$ dependencies identified in the previous stage (Fig. 12) with the results of the parametric finite element calculations (Section 3 and Appendix B) for each thermal cycle to obtain a relationship between the accumulated time $t_{acc}$ and the critical value of the local stress responsible for local microcrack initiation and growth within the TBC, denoted by $\sigma_C$. This stress is regarded as critical because it is associated with the measured values of the damage parameter $D$ via its dependence on $r_c$. The results of the parametric analyses are in the form of curves of the peak values of the stress component which controls TBC failure versus accumulated time for given values of $b/a$ and $\theta_{max}$. By interpolating between these curves with respect to $b/a$, $\theta_{max}$, (assuming $\theta_{min}$ remains always the same), and accumulated time (or TGO thickness), the value of the stress $\sigma_C$ at a time $t_{acc}$ for the corresponding critical value $r_c$ and temperatures $\theta_{max}$ and $\theta_{min}$ can be found.

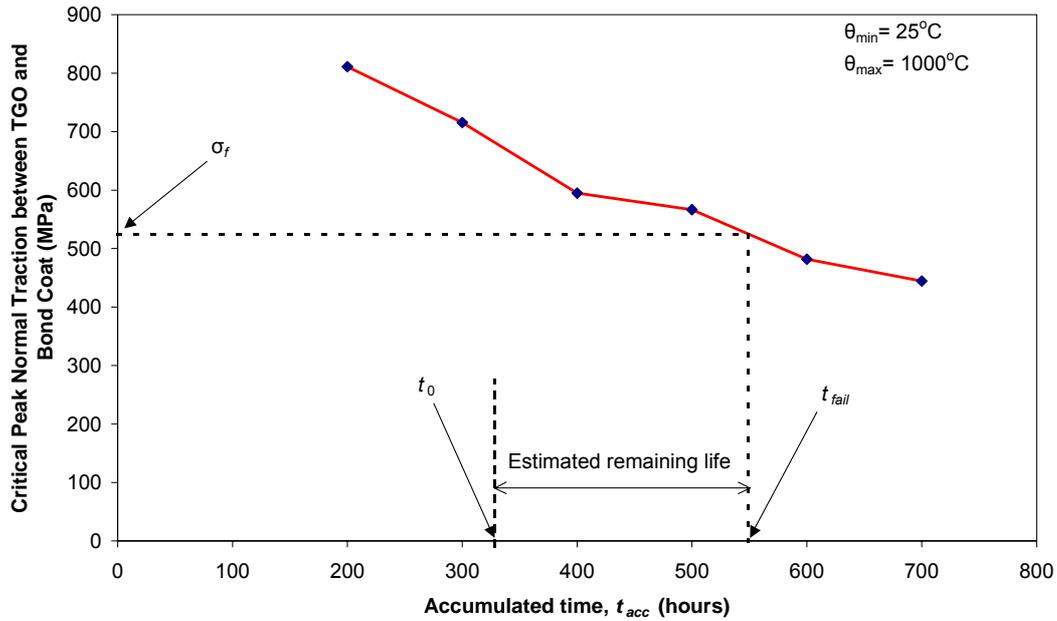

(a)

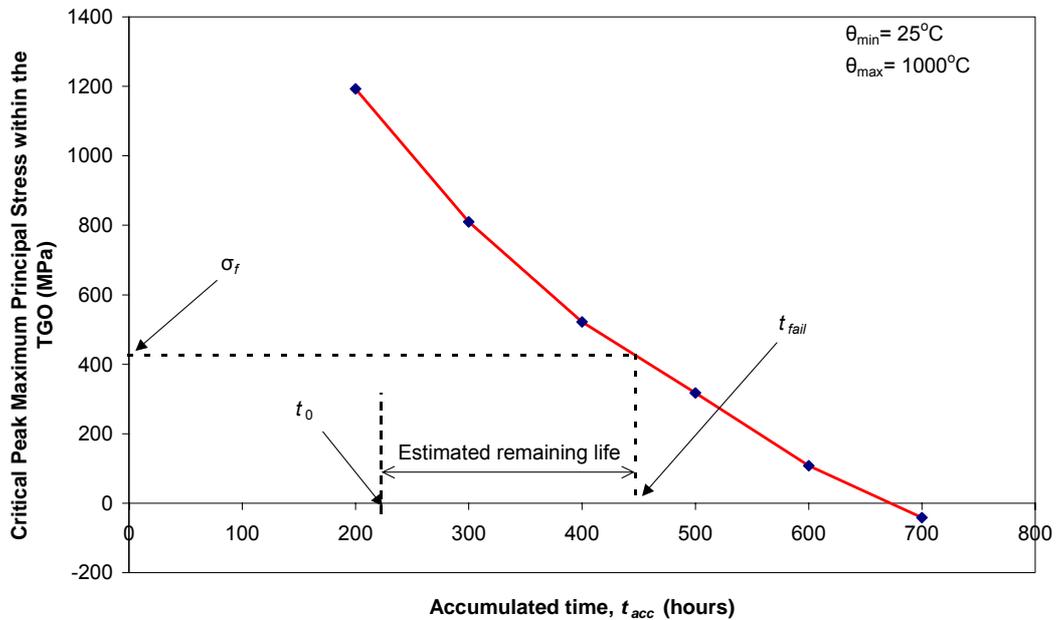

(b)

**Figure 13** Illustration of interpolation to obtain TBC failure time, $t_{fail}$, and remaining lifetime for a coating whose failure is controlled by (a) the peak normal traction on the TGO-bond coat interface and (b) peak maximum principal stress within the TGO.

The life prediction methodology proposed here was numerically implemented and accounts for the following options for the stress interpolation, depending on which stress component controls TBC failure: (i) the peak maximum principal stress within the TGO, (ii) the peak out-of-plane stress in the YSZ top coat, and (iii) the peak interfacial TGO tractions. To illustrate the methodology, and in the absence of conclusive evidence about the dominant failure mode of this TBC system, the results presented here use the peak normal traction, $\sigma_{NT}$, on the TGO-bond coat interface (Fig. 13(a)) and the peak maximum principal stress within the TGO, $\sigma_{MPS}$ (Fig. 13(b)). As mentioned in Section 2.1, the first of these stresses drives the nucleation and propagation of interfacial cracks, and the second drives nucleation and propagation of cohesive cracks within the TGO.

The interpolation process for a chosen thermal profile proceeds as follows:

Step 1: Inputs: Accumulated time $t_{acc}$.

Step: using Eq. 3 for $h$ in terms of $t_{acc}$ and $\theta_{max}$, and values of $r_c$ vs. $t_{acc}$ as in Fig 12, determine the values of TGO thickness ($h_c$) and $r_c$ for the time $t_{acc}$ and the temperature $\theta_{max}$.

Output: values of $h_c$ and $r_c$.

Step 2: Inputs: FE stress results.

Step: For each curve of peak stress vs. accumulated time obtained by FEA, use Eq. 3 and the temperature $\theta$ at which the results were calculated, to transform accumulated time into TGO thickness.

Output: Curves of peak stress vs. $h$ for a set of values of $\theta$ and $b/a$.

Step 3: Inputs: Curves of peak stress vs. $h$ from previous step, value of $h_c$ from Step 1.

Step: Interpolate the data on each curve to obtain the peak stress at a TGO thickness of $h_c$ for the values of $\theta$ and $b/a$ corresponding to that curve.

Output: Values of peak stress at a TGO thickness $h_c$ for a set of values of $\theta$ and $b/a$ (values lie on a 2-d surface).

Step 4: Inputs: Peak stress values from previous step, value of $\theta_{max}$ corresponding to thermal profile associated with measured damage value $D$.

Step: Interpolate between the values from Step 3 with respect to the temperatures used for FE, to obtain the peak stress at a TGO thickness $h_c$ and a temperature $\theta_{max}$ for a range of values of $b/a$.

Output: Values of peak stress at a TGO thickness $h_c$ and a temperature $\theta_{max}$ for a set of values of $b/a$ (values generate a curve).

Step 5: Inputs: Peak stress vs. $b/a$ curve from step 4, value of $r_c$ from Step 1.

Step: Interpolate the curve from Step 5 with respect to $b/a$ to obtain the value of the peak stress corresponding to a TGO thickness of $h_c$, a temperature $\theta_{max}$, and a value of $b/a$ equal to $r_c$. This value is $\sigma_C$.

Output: A value of $\sigma_C$ at a time $t_{acc}$ and a temperature $\theta_{max}$.

These five steps can be repeated for different values of $t_{acc}$ to build up a curve of accumulated time versus critical stress for a given thermal profile. The resulting critical stress versus accumulated time curves for the 25-1000 °C temperature history are shown in Figs. 13(a) and (b). Note that the critical stress $\sigma_C$ is that required to initiate a local crack for the given value of $r_c$ at the given time $t_{acc}$, and is not the stress required to cause total failure at that time.

As was mentioned in Section 2.5, the FE models were run until the TGO thickness reached 3 µm, since it is expected that the coating system is likely to fail by this point. Hence the FE results are only available for TGO thicknesses less than or equal to 3 µm. If some of the values of $t_{acc}$ correspond to TGO thicknesses greater than 3 µm, then the software carries out an extrapolation of the FE results. In some cases this may lead to physically unlikely stress values, and so results must be interpreted with care.

### *4.3 Stage 3: Life Prediction*

In this third and last stage the remaining life of the coating is estimated. The failure criterion used in this stage is based on the understanding that global TBC failure occurs by a cleavage-type mechanism, that is, it is stress, or more strictly speaking energy, driven. The implied assumption here is that failure is initiated from a micro-defect of a given size, requiring a specific stress for the operation of the cleavage mechanism. Under this assumption, for a given temperature cycle, the failure of any given region of the TBC will occur when the local 'driving force' (i.e. stress component which controls TBC failure, denoted as $\sigma_c$ in Section 4.2) reaches a global failure value, $\sigma_f$, independently of all other factors. It is possible to estimate $\sigma_f$ from measurements made on TBCs that have failed, by taking $\sigma_f$ to be the value of $\sigma_c$ corresponding to the last measurement made before failure (see ref. [4] for an example). The global failure value $\sigma_f$ is a property of the material system being examined and is effectively a material property.

The inputs required in this stage are, for each temperature history of interest, the global failure stress $\sigma_f$, the current value of accumulated time, $t_0$, and the curve of $\sigma_c$ versus $t_{acc}$ generated as the output of the previous stage. The procedure to predict the remaining lifetime is:

(i) find the accumulated time from the curve of $\sigma_c$ versus $t_{acc}$ that corresponds to the critical stress for global failure $\sigma_f$ as shown in Fig. 13. This time is the failure time, $t_{fail}$.

(ii) subtract $t_0$ from $t_{fail}$ to obtain the remaining life of the coating.

If the value of $t_0$ is not known, it can be estimated from a damage measurement by interpolating the curve of damage versus accumulated time (shown in Fig. 9). It is possible that more than one value of $t_{fail}$ is obtained from the interpolation in (i).

It should be noted that, in principle, an alternative approach could be followed based on a measured mean stress value rather than the measured damage used so far. It will also rely on an interpolation of the finite element results to estimate an effective roughness ratio ($r_m$) from the measured stress, $\sigma_m$, actual time $t_{acc}$ and the $\theta_{max}$ of interest. However, if the stress is measured using piezo-spectroscopy, the measured stress would, in the most general case, be limited to the mean triaxiality, as discussed in Section 2.2, rather than the component responsible for failure. Thus specific failure modes within the TBC would not be able to be addressed using a measured mean stress approach.

## 5. Conclusions

A mechanism-based life prediction methodology for thermal barrier coating systems has been proposed. It relies on a combination of information about the TBC's morphological characteristics, accumulated TBC damage inferred from non-destructive fluorescence measurements, and numerically predicted local TBC stresses responsible for the initiation of such damage. It incorporates the complex interaction between interfacial and microstructural features, local oxidation mechanisms and time-dependent processes. The methodology is applied to predict the life of an EB-PVD TBC with an MCrAlY bond coat. The results of parametric finite element studies using periodic unit cell techniques revealed the magnitudes of the local TBC stresses known to lead to the failure of this type of coating. The maximum TGO stresses responsible for microcrack nucleation were found to increase with oxidation time and TGO roughness.

The proposed novel procedure to assess TBC failure has been shown to provide a physically-based yet technologically feasible method to predict the onset of TBC damage *in-situ*.

**Appendix A** : TBC Properties

A.1  Elasto-visco-plastic properties of the bond coat and oxide

Table A1: Elastic-plastic material properties of the bond coat and oxide

| Temperature (°C) | Bond coat | | | | TGO | | |
|---|---|---|---|---|---|---|---|
| | E (GPa) | v | α (×10$^6$, K$^{-1}$) | σ$_Y$ (MPa) | E (GPa) | v | α (×10$^6$, K$^{-1}$) |
| 20 | 200 | 0.3 | 13.6 | 426 | 400 | 0.23 | 8.00 |
| 200 | 190 | 0.3 | 14.2 | 412 | 390 | 0.23 | 8.20 |
| 400 | 175 | 0.31 | 14.6 | 396 | 380 | 0.24 | 8.40 |
| 600 | 160 | 0.31 | 15.2 | 362 | 370 | 0.24 | 8.70 |
| 800 | 145 | 0.32 | 16.1 | 284 | 355 | 0.25 | 9.00 |
| 1000 | 120 | 0.33 | 17.2 | 202 | 325 | 0.25 | 9.30 |
| 1100 | 110 | 0.33 | 17.6 | 114 | 320 | 0.25 | 9.60 |

The steady state creep model for the bond coat and TGO is given by

$$\dot{\bar{\varepsilon}}^{cr} = A \ \exp\left(-\frac{Q}{R\theta}\right) \tilde{\sigma}^n \tag{A.1}$$

where $\theta$ is the temperature in K, and $\tilde{\sigma}$ is the uniaxial equivalent Mises stress. The values of the material properties $A$, n and $Q$ for the bond coat and TGO are shown in Table A2.

Table A2: Creep properties of the bond coat and oxide

| | Bond coat | TGO |
|---|---|---|
| n | 4.1 | 2.3 |
| Q (kJ/ (kg mol)) | 263×10$^3$ | 424×10$^3$ |
| A (1/(s MPa$^{-n}$)) | 13.3 | 6805 |

A.2 Effect of sintering time and temperature on the transversely anisotropic elastic moduli of YSZ

The elastic properties of the transversely isotropic YSZ are here fully defined by $E_2$, $E_p/E_2$ (function of sintering time only), and $v_{pp}$ (function of temperature only), where $E_p$ is the isotropic in-plane Young's modulus, $E_2$ is the out-of-plane Young's modulus, and $v_{pp}$ is the Poisson's ratio linking the in-plane directions. The other elastic constants are given by

$$G_p = \frac{E_p}{2(1+v_{pp})}, \quad v_{2p} = v_{p2}\frac{E_2}{E_p}, \quad G_2 = G_p\frac{E_p}{E_2}, \quad v_{p2} = v_{pp}\frac{E_p}{E_2}. \tag{A.2}$$

The values of $E_2$, $E_p/E_2$, and $v_{pp}$ are given in Tables A1, A2 and A3, respectively.

Table A3: $E_2$ (GPa) as a function of temperature and sintering time

| Time (hours) | Temperature (°C) | | | | | | |
|---|---|---|---|---|---|---|---|
| | 25 | 200 | 400 | 600 | 800 | 1000 | 1200 |
| 0 | 183 | 175 | 168 | 163 | 159 | 155 | 153 |
| 20 | 196 | 187 | 179 | 174 | 169 | 166 | 164 |
| 60 | 203 | 194 | 186 | 180 | 176 | 172 | 170 |
| 100 | 205 | 196 | 188 | 182 | 178 | 174 | 171 |
| 200 | 207 | 198 | 190 | 184 | 179 | 176 | 173 |
| 400 | 209 | 199 | 191 | 185 | 180 | 177 | 174 |
| 600 | 209 | 200 | 192 | 186 | 181 | 177 | 175 |

Table A4: $E_p/E_2$ as a function of sintering time

| Time (hours) | 0 | 20 | 60 | 100 | 200 | 400 | 600 |
|---|---|---|---|---|---|---|---|
| $E_p/E_2$ | 0.01 | 0.169 | 0.258 | 0.281 | 0.303 | 0.318 | 0.325 |

Table A5: $v_{pp}$ as a function of temperature

| Temperature (°C) | 25 | 200 | 400 | 600 | 800 | 1000 | 1200 |
|---|---|---|---|---|---|---|---|
| $v_{pp}$ | 0.1 | 0.1 | 0.1 | 0.11 | 0.11 | 0.12 | 0.12 |

**Appendix B**: Summary of the predicted effects of the geometric parameters and peak cycle temperature on the stress maxima relevant to the different TBC failure modes at room temperature.

| b/a | b (μm) | Temp. (°C) | Acc. Time (h) | TGO thickness (μm) | Max. princ. stress in TGO (MPa) | Max. $\sigma_{22}$ stress in YSZ (MPa) | Max. normal TGO-YSZ traction (MPa) | Max. normal TGO-BC traction (MPa) |
|---|---|---|---|---|---|---|---|---|
| 0.196 | 0.79 | 1000 | 0 | 0.00 | 0 | 129 | 125 | 125 |
| 0.196 | 0.79 | 1000 | 77 | 2.05 | 1015 | 1016 | 811 | 986 |
| 0.196 | 0.79 | 1000 | 193 | 2.78 | 1214 | 1204 | 846 | 1137 |
| 0.196 | 0.79 | 1000 | 241 | 3.00 | 1152 | 1143 | 786 | 1064 |
| 0.196 | 0.79 | 1100 | 0 | 0.00 | 0 | 87 | 84 | 84 |
| 0.196 | 0.79 | 1100 | 11 | 1.82 | 896 | 787 | 668 | 701 |
| 0.196 | 0.79 | 1100 | 38 | 2.76 | 988 | 950 | 898 | 927 |
| 0.196 | 0.79 | 1100 | 49 | 3.00 | 960 | 954 | 717 | 934 |
| 0.323 | 2.47 | 1000 | 0 | 0.00 | 0 | 93 | 92 | 92 |
| 0.323 | 2.47 | 1000 | 61 | 1.91 | 1178 | 907 | 684 | 670 |
| 0.323 | 2.47 | 1000 | 167 | 2.65 | 1143 | 1135 | 786 | 990 |
| 0.323 | 2.47 | 1000 | 241 | 3.00 | 931 | 1031 | 688 | 909 |
| 0.323 | 2.47 | 1100 | 0 | 0.00 | 0 | 95 | 87 | 87 |
| 0.323 | 2.47 | 1100 | 8 | 1.62 | 515 | 684 | 468 | 511 |
| 0.323 | 2.47 | 1100 | 27 | 2.46 | 1050 | 1087 | 678 | 872 |
| 0.323 | 2.47 | 1100 | 49 | 3.00 | 922 | 1052 | 630 | 922 |
| 0.479 | 4.54 | 1000 | 0 | 0.00 | 0 | 161 | 141 | 141 |
| 0.479 | 4.54 | 1000 | 49 | 1.77 | 1239 | 904 | 493 | 696 |
| 0.479 | 4.54 | 1000 | 145 | 2.53 | 1479 | 1183 | 630 | 895 |
| 0.479 | 4.54 | 1000 | 241 | 3.00 | 1033 | 1097 | 565 | 740 |
| 0.479 | 4.54 | 1100 | 0 | 0.00 | 0 | 134 | 134 | 134 |
| 0.479 | 4.54 | 1100 | 6 | 1.48 | 559 | 420 | 392 | 411 |
| 0.479 | 4.54 | 1100 | 20 | 2.22 | 1454 | 673 | 539 | 776 |
| 0.479 | 4.54 | 1100 | 41 | 2.82 | 1405 | 834 | 598 | 881 |
| 0.479 | 4.54 | 1100 | 49 | 3.00 | 1415 | 843 | 607 | 1007 |